\documentclass[twocolumn, 11pt]{article}
\usepackage{amssymb, latexsym, amsmath,graphicx, subfigure}

\begin{document}





\noindent
\textbf{On the Swimming of \textit{Dictyostelium} amoebae}

The conventional mode for amoeboid locomotion is crawling.  N. P. Barry and M. S. Bretscher recently demonstrated that \textit{Dictostelium} amoebae are also capable of swimming towards
chemoattractants \cite{Barry2010}.  They hypothesized that the mechanism for
swimming is intimately related to crawling.  When crawling, the cell front
bifurcates, and protrusions move backwards, relative to the cell.  The authors conjecture that floating cells executing these same motions will swim.  In this letter, we show that, indeed, the shape changes of a crawling cell are sufficient for swimming.

\begin{figure}[bt]
	\begin{center}
	\includegraphics[scale=0.9]{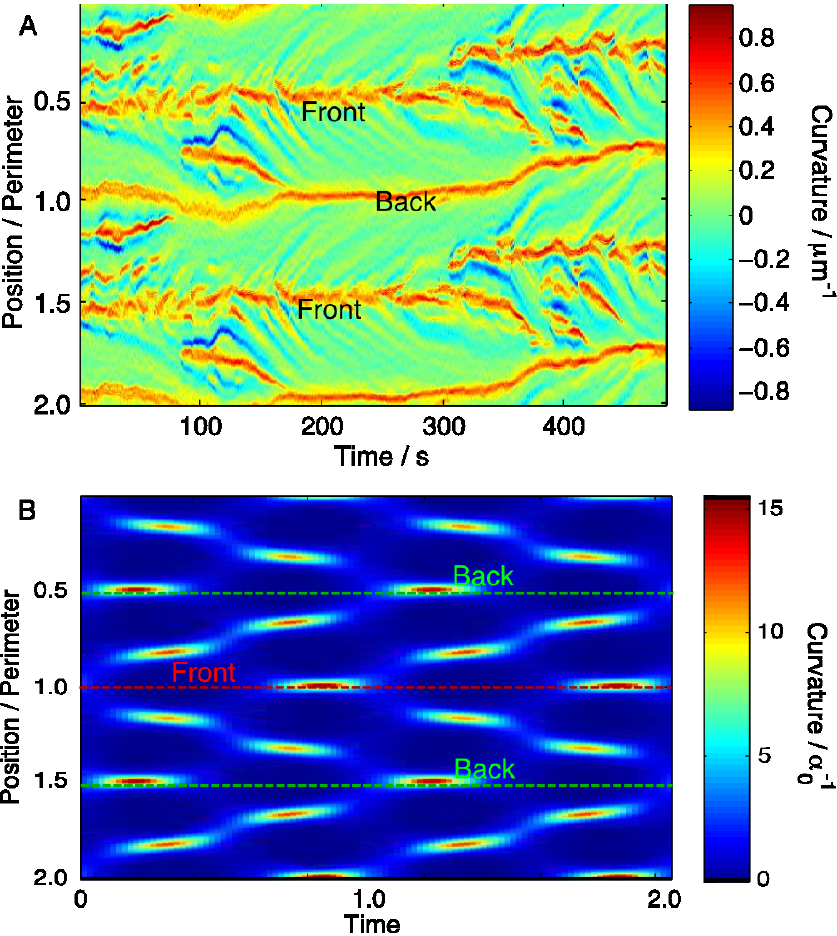}
	\end{center}
		\caption{Curvature space-time plots for the contours of (A) a typical crawling
		cell and (B) the Shapere-Wilczek swimmer.  To prevent a loss of detail at the edges,
		the curvature has been plotted over two contour lengths.  Note the herringbone
		structure---regions of high curvature bifurcate at the front and travel towards the back.}
		\label{fig:curvature}
\end{figure}

To obtain cell geometries, we flattened developed \textit{D. disoideum} (cytosolic GFP in AX2) in an $\sim$4$\mu$m tall chamber \cite{West2010}.  Cells were imaged using confocal microscopy ($f_{image}$=1/s) for 500($\pm$140) s.  From these images, cell contours were retrieved \cite{Debreuve2010}.

The contour curvature plot, \ref{fig:curvature}A, shows that, indeed, protrusions move from the cell front towards the back.  This behavior is also seen in the two-dimensional low Reynolds number swimmer of A. Shapere and F. Wilczek \cite{Shapere1989}.  Therefore, crawling motion seems consistent with swimming.

To more rigorously evaluate this claim, we solved the Stokes flow with no-slip boundary conditions at the cell, and an open boundary, zero normal stress, a distance 250 $\mu$m away.  By calculating the appropriate counterflow (cf. \cite{Shapere1989}), we were able to determine the translational and rotational velocities for the cell, had the cell not been attached to a substrate.

\begin{figure}[htb]
	\begin{center}
	\includegraphics[scale=0.9]{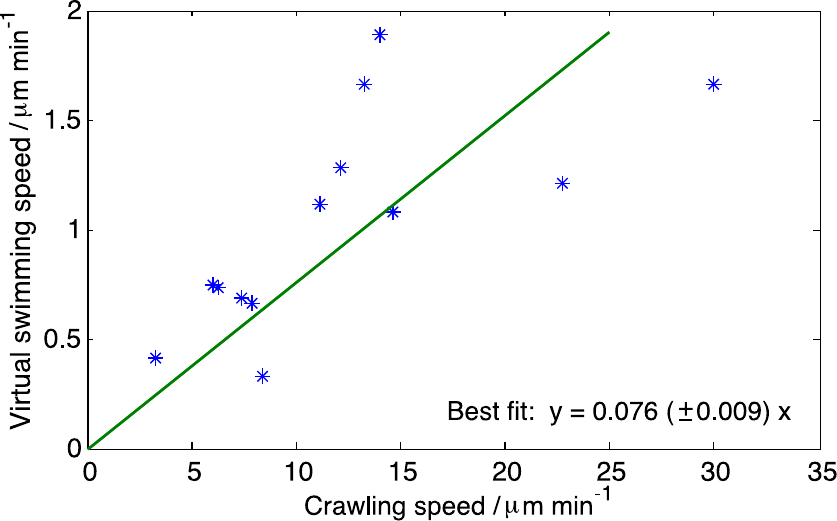}
	\end{center}
		\caption{The crawling and virtual swimming velocities along the direction
		of polarization.  Constrained least squares regression shows that the cells crawl 12 times
		faster than they swim.}
		\label{fig:swimVsCrawl}
\end{figure}

We analyzed the virtual swimming velocity of $n=13$ cells, and found that, for all cells, the time-averaged component along the direction of polarization was positive.\footnote{Defined as the direction from the cell's tail to its centroid.}  The average directed swimming speed was 1.0($\pm$0.5) $\mu$m/min.  For crawling, this was found to be 12($\pm$7) $\mu$m/min.  The correlation coefficient $r=0.6943$ with $p=0.009$.  Therefore, cells that crawl faster are also faster swimmers (figure \ref{fig:swimVsCrawl}).

Despite the qualitative agreement with \cite{Barry2010}, there are discrepancies.  In \cite{Barry2010}, the cells swim at 4.2 $\mu$/min and crawl at 3.8 $\mu$m/min.  Our crawling speed is 3 times higher, and our virtual swimming speed is 4 times lower. On the one hand, flattened cells are less active in producing pseudopodia (data not shown) and thus we expect the virtual swimmer to be slower. On the other hand,  when confined between two plates, a cell may migrate more efficiently: a cell crawling without a ceiling is likely to produce a pseudopod away from the substrate, and only when this pseudopod contacts the substrate can produce propulsion.  We expect the result to hold for neutrophils by analogy.

\noindent
\textbf{Albert J. Bae and Eberhard Bodenschatz} \\
Max Planck Institute for Dynamics and Self-Organization,\\Am Fassberg 17, 37077 G\"{o}ttingen, Germany.\\
Laboratory of Atomic and Solid State Physics, Department of Physics,\\
Cornell University, Ithaca, NY 14853, USA. \\
Institute for Nonlinear Dynamics, \\
University of G\"ottingen, Germany.\\

\end{document}